\documentclass[aps,prb,twocolumn,superscriptaddress,floatfix,longbibliography,raggedbottom]{revtex4-2}
\usepackage{amsmath,amssymb,amsthm}
\usepackage{physics}
\usepackage{amsfonts}
\usepackage{mathrsfs}
\usepackage{graphicx}
\usepackage{tabularx}
\usepackage{enumerate}
\usepackage{dcolumn}
\usepackage{bm}
\usepackage{xcolor}
\usepackage[colorlinks,linkcolor=blue,citecolor=blue,urlcolor=blue]{hyperref}

\graphicspath{{./}}

\begin{document}

\title{Non-Hermitian quantum phase transitions in the XY model induced by staggered imaginary Dzyaloshinskii–Moriya interaction}

\author{Hong Jiang}
\affiliation{Center of Materials Science and Optoelectronics Engineering, College of Materials Science and Opto-Electronic Technology, University of Chinese Academy of Sciences, Beijing 100049, China}

\author{Xiang-Ping Jiang}
\email{2015iopjxp@gmail.com}
\affiliation{School of Physics, Hangzhou Normal University, Hangzhou, Zhejiang 311121, China}

\author{Yan-Chao Li}
\email{ycli@ucas.ac.cn}
\affiliation{Center of Materials Science and Optoelectronics Engineering, College of Materials Science and Opto-Electronic Technology, University of Chinese Academy of Sciences, Beijing 100049, China}

\begin{abstract}
We investigate non-Hermitian quantum phase transitions driven by staggered imaginary Dzyaloshinskii–Moriya (DM) interactions in a transverse-field $XY$ chain. By virtue of a staggered nonunitary transformation, we exactly diagonalize the Hamiltonian. For $D<1$, this procedure maps the non-Hermitian model onto a standard Hermitian $XY$ chain, allowing analytical derivation of the full phase boundaries. The parameter line $D=1$ forms an exceptional boundary with coalesced quasiparticle eigenvalues and eigenvectors, while the entire $D>1$ regime falls into the $\mathcal{RT}$-symmetry-broken phase containing two distinct $z$-ferromagnetic phases. We examine common quantum-information probes and show that conventional measures, such as entanglement entropy, quantum discord, and quantum coherence, fail to capture the $\mathcal{RT}$ symmetry-breaking transition. To address this deficiency, we propose a novel coherence measure $\widetilde{\mathrm{QC}}_{\max}^{\rm LR}$ based on $\mathcal{RT}$ symmetry and complex-conjugate eigenpair correlations. Embedding intrinsic non-Hermitian information of left and right eigenstates, $\widetilde{\mathrm{QC}}_{\max}^{\rm LR}$ reliably identifies the $\mathcal{RT}$ symmetry-breaking transition and accurately resolves all magnetic phase boundaries.

\end{abstract}

\maketitle

\section{Introduction}
\label{sec:introduction}

The spin-$1/2$ $XY$ chain is a standard exactly solvable model for studying quantum phase transitions~\cite{JordanWigner1928,Lieb1961,BarouchMcCoyDresden1970,Barouch1971}.  The Jordan--Wigner transformation maps it to free fermions, allowing its energy spectrum, spin correlations, and phase boundaries to be obtained directly.  The model contains ferromagnetic (FM), gapless, and paramagnetic (PM) phases and has been widely used to study quantum critical behavior.  Non-Hermitian $XY$ chains are obtained when the exchange or the magnetic field becomes complex~\cite{Ashida2020,DingFangMa2022,MedenGrunwaldKennes2023}.  Their energies can be complex, their left and right eigenstates are different, and their eigenvalues and eigenvectors can merge at exceptional points~\cite{Bender1998,Heiss2012,Bergholtz2021,Wiersig2023}.  When the exchange anisotropy term $\gamma$ takes an imaginary value, the model possesses rotation–time-reversal  ($\mathcal{RT}$) symmetry~\cite{ZhangSong2013}. Earlier studies have identified its exceptional boundary and critical behaviors~\cite{ZhangSong2013Geometric,LiuBatchelor2025,HenryLiuBatchelor2025,LuoMeden2026,LiLiuBatchelor2026}. Complementary investigations have explored the phase diagrams, correlation properties, and dynamical behaviors of non-Hermitian $XY$ chains with uniform, staggered, and complex fields~\cite{LiuXuLi2021,PiLu2021,LakkarajuSenDe2021,AgarwalEtAl2024,WangXu2026}.

The Dzyaloshinskii--Moriya (DM) interaction is an antisymmetric exchange generated by spin--orbit coupling when inversion symmetry is broken~\cite{Dzyaloshinsky1958,Moriya1960,KatsuraNagaosaBalatsky2005,YangLiangCui2023}.  A uniform DM interaction makes the fermion dispersion asymmetric and supports a gapless chiral phase.  Recent work introduced this interaction into a non-Hermitian $XY$ chain and showed that it breaks $\mathcal{RT}$ symmetry and produces chiral phases~\cite{ZhangDM2026}.  For a staggered DM interaction, the two spin sublattices rotate in opposite directions.  For real staggered DM couplings, the spatially varying phase factor can be completely gauged out through site-dependent spin transformations, rendering the model analytically equivalent to a conventional Hermitian system~\cite{PerkCapel1976,ShekhtmanEntinWohlmanAharony1992,Baran2018}. Benefiting from this Hermitian reduction, existing studies have extensively investigated the critical phenomena, entanglement characteristics, and magnetic ordering behaviors of Hermitian spin chains hosting real staggered DM interactions~\cite{Ma2011,Su2022,LiuStaggered2024,Sun2024}. However, the situation changes fundamentally when the staggered DM coupling becomes imaginary, which generates unequal bipartite spin-exchange amplitudes and induces inherent non-Hermiticity. Although the physical effects of such asymmetric exchange amplitudes have been briefly discussed in non-Hermitian $XXZ$ chains~\cite{Ghosh2010}, the corresponding phase diagrams, symmetry-breaking features, and quantum phase transitions in imaginary-staggered-DM $XY$ chains still lack systematic exploration.

Quantum information probes have become effective tools for characterizing quantum phase transitions, overcoming the inherent limitation of conventional methods that rely on prior symmetry information of the system. Owing to their simplicity and robustness, these probes are widely recognized as ideal candidates for exploring critical behaviors. Various quantum information measures have been extensively applied to identify phase transitions in Hermitian systems, covering topological transitions and Berezinskii–Kosterlitz–Thouless transitions~\cite{Osterloh2002,Peschel2003,Calabrese2004,PeschelEisler2009,EislerPeschel2010,KarpatCakmakFanchini2014,Lv2022,SuEtAl2024,LiEtAl2024Multipartite,Ollivier2001,Henderson2001,Sarandy2009,BaumgratzCramerPlenio2014,Girolami2014}. These probes have recently been extended to non-Hermitian systems to resolve exceptional boundaries and quantum phase transitions~\cite{Herviou2019,Chen2021,TzengEtAl2021,PiresMacri2021,Miao2024,ZhangYu2025,AgarwalEtAl2026}. However, related studies are still in the preliminary stage, and the universal applicability of quantum information probes in non-Hermitian phase transition characterization remains to be systematically validated.

Specifically, the non-Hermitian $XY$ chain with staggered imaginary DM interactions retains intrinsic $\mathcal{RT}$ symmetry. The spontaneous $\mathcal{RT}$ symmetry breaking and the corresponding phase transition behaviors in the symmetry-broken phase remain unclear. Furthermore, the feasibility of conventional quantum information quantities for diagnosing non-Hermitian phase transitions in this model lacks systematic investigation. Different from Hermitian systems, quantum information metrics in non-Hermitian systems possess two rigorous definitions: the self-normalized definition constructed merely from right eigenstates, and the biorthonormal definition integrating both left and right eigenstates. The optimal definition that guarantees higher accuracy and universality for non-Hermitian phase transition detection currently lacks conclusive justification.

In this work, we determine the phase diagram of the $XY$ chain with a staggered imaginary DM interaction in a transverse field and use quantum-information quantities to probe its phase transitions.  We construct a staggered nonunitary rotation to obtain the analytic phase boundaries and determine the magnetic phases from spin correlations and the quasiparticle spectrum.  We then compare self-normal and biorthogonal entanglement entropy, QD, and quantum coherence against these boundaries.  Our results reveal a first-order transition inside the $\mathcal{RT}$-broken phase.  For the biorthogonal coherence, we separate the real and imaginary parts of the coherence matrix and show that the balance between the corresponding coherence contributions causes two eigenvalues of the full complex matrix to coalesce, identifying the exceptional boundary.

The rest of this paper is organized as follows. Section~\ref{sec:model} derives the mapping and the phase diagram.  Section~\ref{sec:information} presents the quantum-information results, and Sec.~\ref{sec:conclusion} gives the conclusions.

\section{Model and phase diagram}
\label{sec:model}

\begin{figure*}[t]
\centering
\includegraphics[width=\textwidth]{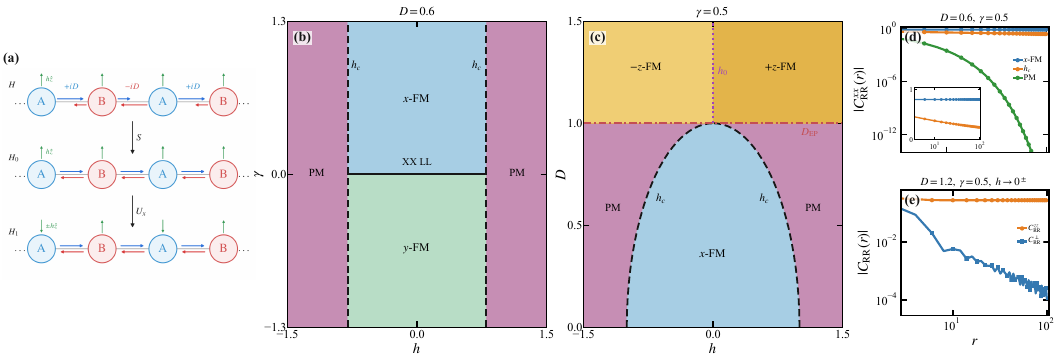}
\caption{Model, magnetic phase diagrams, and spin correlations at $J=1$. (a) The unequal spin-exchange amplitudes in $H$ and the transformations to $H_0$ and $H_1$ by $\mathcal S$ and $\mathcal U_x$. (b) $(h,\gamma)$ phase diagram at $D=0.6$. (c) $(h,D)$ phase diagram at $\gamma=0.5$. The black dashed, purple dotted, and red dash-dotted lines mark $h=\pm h_c$, $h=h_0$, and $D=D_{\rm EP}$, respectively. (d) $|C_{\rm RR}^{xx}(r)|$ in the $x$-FM phase, at $h=h_c$, and in the PM phase for $D=0.6$ and $\gamma=0.5$. (e) $|C_{\rm RR}^{zz}(r)|$ and $C_{\rm RR}^{\perp}(r)$ for $D=1.2$, $\gamma=0.5$, and $h\rightarrow0^\pm$.}
\label{fig:model_phase}
\end{figure*}

\subsection{Nonunitary mapping and magnetic phases}

We consider the spin-$1/2$ $XY$ chain with staggered imaginary DM interaction in a transverse field, described by the following Hamiltonian:
\begin{align}
H={}&-\frac{J}{2}\sum_{j=1}^{N}\Big[
(1+\gamma)\sigma_j^x\sigma_{j+1}^x
+(1-\gamma)\sigma_j^y\sigma_{j+1}^y \nonumber\\
&+\mathrm{i}(-1)^{j-1}D
\left(\sigma_j^x\sigma_{j+1}^y
-\sigma_j^y\sigma_{j+1}^x\right)\Big]
-h\sum_{j=1}^{N}\sigma_j^z.
\label{eq:model}
\end{align}
Here $J>0$, $h$ is the transverse field, $\gamma$ is the exchange anisotropy, and $D$ denotes the staggered imaginary DM interaction.  Without loss of generality, we take $D\geq0$.

Using $\sigma_j^{\pm}=(\sigma_j^x\pm\mathrm{i}\sigma_j^y)/2$, Eq.~\eqref{eq:model} becomes
\begin{align}
H={}&-J\sum_j\Big\{
\left[1-(-1)^{j-1}D\right]\sigma_j^+\sigma_{j+1}^-
\nonumber\\
&+\left[1+(-1)^{j-1}D\right]\sigma_j^-\sigma_{j+1}^+
\nonumber\\
&+\gamma\left(\sigma_j^+\sigma_{j+1}^+
+\sigma_j^-\sigma_{j+1}^-\right)\Big\}
-h\sum_j\sigma_j^z.
\label{eq:raising_lowering}
\end{align}
The imaginary DM interaction enters the two spin-exchange amplitudes with opposite signs, so that the two amplitudes become unequal.  Previous work has removed a similar exchange asymmetry in non-Hermitian $XXZ$ chains by introducing a local nonunitary similarity transformation~\cite{Ghosh2010}.  Following this idea, we construct a staggered nonunitary rotation for this $XY$ chain.  For $D\neq1$,
\begin{equation}
\mathcal S=\prod_{j=1}^{N}
\exp\!\left[
\frac{(-1)^{j-1}}{8}
\ln\!\left(\frac{1+D}{1-D}\right)\sigma_j^z
\right].
\label{eq:similarity}
\end{equation}
It rescales the raising and lowering operators according to
\begin{equation}
\mathcal S\sigma_j^{\pm}\mathcal S^{-1}
=\left(\frac{1+D}{1-D}\right)^{\pm(-1)^{j-1}/4}
\sigma_j^{\pm}.
\label{eq:S_action}
\end{equation}
The rotation makes the two spin-exchange amplitudes in Eq.~\eqref{eq:raising_lowering} equal to $\sqrt{1-D^2}$ and transforms the Hamiltonian into
\begin{align}
H_0\equiv\mathcal S H\mathcal S^{-1}
={}&-\frac{J}{2}\sum_j\Big[
\left(\sqrt{1-D^2}+\gamma\right)\sigma_j^x\sigma_{j+1}^x
\nonumber\\
&+\left(\sqrt{1-D^2}-\gamma\right)\sigma_j^y\sigma_{j+1}^y\Big]
-h\sum_j\sigma_j^z.
\label{eq:H0}
\end{align}

For $D<1$, $\sqrt{1-D^2}$ is real.  We define
\begin{equation}
J_{\rm eff}=J\sqrt{1-D^2},\qquad
\gamma_{\rm eff}=\frac{\gamma}{\sqrt{1-D^2}}.
\label{eq:effective_parameters}
\end{equation}
With these effective parameters, the staggered imaginary DM interaction is absorbed into $J_{\rm eff}$ and $\gamma_{\rm eff}$, and Eq.~\eqref{eq:H0} takes the standard form
\begin{align}
H_0={}&-\frac{J_{\rm eff}}{2}\sum_j\Big[
(1+\gamma_{\rm eff})\sigma_j^x\sigma_{j+1}^x
+(1-\gamma_{\rm eff})\sigma_j^y\sigma_{j+1}^y\Big]
\nonumber\\
&-h\sum_j\sigma_j^z.
\label{eq:Hermitian_XY}
\end{align}
The transformed Hamiltonian is therefore a Hermitian $XY$ chain.  Its phase diagram has been established in Refs.~\cite{Lieb1961,BarouchMcCoyDresden1970,Barouch1971}.  Using this mapping, we obtain the magnetic boundary
\begin{equation}
h_c=J_{\rm eff}=J\sqrt{1-D^2},\qquad D<1.
\label{eq:hc}
\end{equation}
Inside $|h|<h_c$, positive and negative $\gamma$ give the $x$-FM and $y$-FM phases, respectively.  At $\gamma=0$, the model reduces to the gapless $XX$ chain, whose ground state is a Luttinger liquid (LL).  For $|h|>h_c$, the system is paramagnetic.  These phases form the phase diagram in Fig.~\ref{fig:model_phase}(b) and the $D<1$ part of Fig.~\ref{fig:model_phase}(c).

To confirm the phase diagram obtained from the exact mapping, we calculate the right-state spin correlations of the Hamiltonian in Eq.~\eqref{eq:model},
\begin{equation}
C_{\rm RR}^{\alpha\alpha}(r)=
\frac{\langle R_0|\sigma_j^\alpha\sigma_{j+r}^\alpha|R_0\rangle}
{\langle R_0|R_0\rangle},
\qquad \alpha=x,y,z.
\label{eq:spin_correlations}
\end{equation}
Here $|R_0\rangle$ denotes the right ground state. The correlation magnitudes shown in Figs.~\ref{fig:model_phase}(d) and~\ref{fig:model_phase}(e) are averaged over the two sites in a unit cell. For $D=0.6$ and $\gamma=0.5$, Fig.~\ref{fig:model_phase}(d) shows that $C_{\rm RR}^{xx}(r)$ approaches a finite value in the $x$-FM phase, decays algebraically at $h=h_c=0.8$, and decays exponentially in the paramagnetic phase. These characteristic behaviors identify the two magnetic phases and confirm that their boundary lies at the value predicted by Eq.~\eqref{eq:hc}.

We next consider the region $D>1$.  In this region, $\sqrt{1-D^2}=\mathrm{i}\sqrt{D^2-1}$, and Eq.~\eqref{eq:H0} becomes
\begin{align}
H_0={}&-\frac{J}{2}\sum_j\Big[
\left(\gamma+\mathrm{i}\sqrt{D^2-1}\right)\sigma_j^x\sigma_{j+1}^x
\nonumber\\
&\hspace{1.15cm}-\left(\gamma-\mathrm{i}\sqrt{D^2-1}\right)
\sigma_j^y\sigma_{j+1}^y\Big]
-h\sum_j\sigma_j^z.
\label{eq:H0_strong}
\end{align}
The two transverse-exchange terms now carry opposite overall signs.  To write this Hamiltonian in the standard form of a non-Hermitian $XY$ chain with imaginary anisotropy, we further rotate every odd spin by $\pi$ about the $x$ axis,
\begin{equation}
\mathcal U_x=\prod_{j\,\mathrm{odd}}\sigma_j^x.
\label{eq:Ux}
\end{equation}
This rotation leaves $\sigma_j^x\sigma_{j+1}^x$ unchanged and changes the sign of $\sigma_j^y\sigma_{j+1}^y$, mapping the exchange part of $H_0$ to the standard non-Hermitian $XY$ form.  But it also changes the sign of $\sigma_j^z$ on the odd sites, so that the uniform field becomes a staggered field.  We therefore obtain
\begin{align}
H_1&\equiv\mathcal U_xH_0\mathcal U_x^{-1}\nonumber\\
&=-\frac{J}{2}\sum_j\Big[
\left(\gamma+\mathrm{i}\sqrt{D^2-1}\right)\sigma_j^x\sigma_{j+1}^x
\nonumber\\
&\hspace{1.15cm}+\left(\gamma-\mathrm{i}\sqrt{D^2-1}\right)
\sigma_j^y\sigma_{j+1}^y\Big]\nonumber\\
&\quad+h\sum_j(-1)^{j-1}\sigma_j^z.
\label{eq:H1}
\end{align}
Equation~\eqref{eq:H1} describes a non-Hermitian $XY$ chain with imaginary anisotropy and a staggered field. The odd-site rotation maps the uniform $z$ magnetization of the original chain onto a staggered magnetization in $H_1$. To determine the magnetic order for $D>1$, we calculate the longitudinal correlation $C_{\rm RR}^{zz}(r)$ and the overall transverse correlation in the $xy$ plane,
\begin{equation}
C_{\rm RR}^{\perp}(r)=
\left[
\frac{|C_{\rm RR}^{xx}(r)|^2+|C_{\rm RR}^{yy}(r)|^2}{2}
\right]^{1/2}.
\label{eq:Cperp}
\end{equation}
For $D=1.2$ and $\gamma=0.5$, Fig.~\ref{fig:model_phase}(e) shows that $|C_{\rm RR}^{zz}(r)|$ approaches a finite value as $h\rightarrow0^\pm$, while $C_{\rm RR}^{\perp}(r)$ decays to zero. These correlations establish long-range ferromagnetic order along the $z$ direction. As shown in Fig.~\ref{fig:spectrum_transition}(d), the longitudinal magnetization is positive for $h>0$ and negative for $h<0$. We therefore identify the two regions as the $+z$-FM and $-z$-FM phases. They meet at $h=h_0=0$. The spectrum and ground-state energy in the next subsection establish this boundary as a first-order transition.

\subsection{Quasiparticle spectrum and the \texorpdfstring{$\mathcal{RT}$}{RT} transition}
\begin{figure}[!t]
	\centering
	\includegraphics[width=\columnwidth]{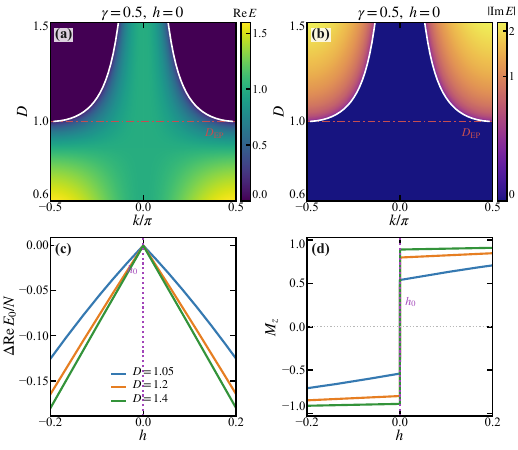}
	\caption{Quasiparticle spectrum and the first-order transition at $J=1$. (a) $\operatorname{Re}E(k)$ and (b) $|\operatorname{Im}E(k)|$ at $\gamma=0.5$ and $h=h_0$. The red dash-dotted lines mark $D=D_{\rm EP}$, and the white curves give the exceptional momenta in Eq.~\eqref{eq:kEP}. (c) Change in the real ground-state energy for $D>1$. (d) Longitudinal magnetization for the same parameters. The purple dotted lines in (c) and (d) mark $h=h_0$.}
	\label{fig:spectrum_transition}
\end{figure}

Earlier studies established the rotation--time-reversal symmetry of non-Hermitian $XY$ chains with uniform and staggered transverse fields~\cite{LakkarajuSenDe2021,AgarwalEtAl2024}.  The mapped Hamiltonian $H_1$ in Eq.~\eqref{eq:H1} is therefore invariant under $\mathcal{RT}$,
\begin{equation}
\begin{gathered}
\mathcal R=\exp\!\left(-\frac{\mathrm{i}\pi}{4}\sum_j\sigma_j^z\right),\qquad
\mathcal T\,\mathrm{i}\,\mathcal T^{-1}=-\mathrm{i},\\[2pt]
(\mathcal R\mathcal T)H_1(\mathcal R\mathcal T)^{-1}=H_1.
\end{gathered}
\label{eq:RT_symmetry}
\end{equation}
The transformations $\mathcal S$ and $\mathcal U_x$ in Eqs.~\eqref{eq:similarity} and~\eqref{eq:Ux} preserve this $\mathcal{RT}$ symmetry.  Therefore, the original Hamiltonian $H$ in Eq.~\eqref{eq:model} also has $\mathcal{RT}$ symmetry, written in the original spin basis as
\begin{equation}
	\begin{gathered}
	\left[\mathcal S^{-1}\mathcal U_x^{-1}(\mathcal R\mathcal T)\mathcal U_x\mathcal S\right]H\left[\mathcal S^{-1}\mathcal U_x^{-1}(\mathcal R\mathcal T)\mathcal U_x\mathcal S\right]^{-1}=H.
	\label{eq:RT_oneline}
	\end{gathered}
\end{equation}
To further examine the non-Hermitian properties of this model, we derive the quasiparticle spectrum directly from Eq.~\eqref{eq:model} and use it to identify the location of the $\mathcal{RT}$-symmetry-breaking transition.  After applying the odd-site rotation $\mathcal U_x$ to $H$ and performing the Jordan--Wigner transformation~\cite{JordanWigner1928}, in the basis $\Psi_k=(c_k,c_{-k}^{\dagger},c_{k+\pi},c_{-k-\pi}^{\dagger})^{\mathsf T}$, with $-\pi/2<k\leq\pi/2$, the Bogoliubov--de Gennes matrix is
\begin{equation}
\resizebox{0.98\columnwidth}{!}{$\displaystyle
\mathcal H_{\rm BdG}(k)=2\begin{pmatrix}
J\gamma\cos k&-\mathrm{i} J(1+D)\sin k&-h&0\\
\mathrm{i} J(1-D)\sin k&-J\gamma\cos k&0&h\\
-h&0&-J\gamma\cos k&\mathrm{i} J(1+D)\sin k\\
0&h&-\mathrm{i} J(1-D)\sin k&J\gamma\cos k
\end{pmatrix}$}.
\label{eq:BdG}
\end{equation}
Its characteristic polynomial factorizes as
\begin{align}
\det\!\left[E-\mathcal H_{\rm BdG}(k)\right]
=\prod_{\nu=\pm1}\Big\{E^2-4J^2\gamma^2\cos^2k
\nonumber\\[-2pt]
-4\left[J\sqrt{1-D^2}\sin k+\nu h\right]^2\Big\},
\label{eq:charpoly}
\end{align}
which gives four quasiparticle bands,
\begin{equation}
E(k)=\pm2\sqrt{J^2\gamma^2\cos^2 k+\left[J\sqrt{1-D^2}\sin k\pm h\right]^2}.
\label{eq:spectrum}
\end{equation}
The two signs are chosen independently.
For $D<1$, the spectrum is real and closes at the magnetic boundary $|h|=h_c$ given by Eq.~\eqref{eq:hc}.  At $\gamma=0$, it is gapless throughout $|h|<h_c$.  The spectrum therefore reproduces the Hermitian phase diagram obtained from the mapping.

At $D=1$, the characteristic polynomial becomes
\begin{equation}
\det\!\left[E-\mathcal H_{\rm BdG}(k)\right]
=\left[E^2-4J^2\gamma^2\cos^2k-4h^2\right]^2.
\label{eq:D1_charpoly}
\end{equation}
Equation~\eqref{eq:D1_charpoly} shows that the four quasiparticle bands merge pairwise at $D=1$.  At $k=\pi/2$, the eigenvalues and eigenvectors in each pair coalesce, forming two second-order exceptional points at $E=\pm2h$.  These two exceptional points meet at zero energy when $h=h_0$.  The simultaneous coalescence of eigenvalues and eigenvectors is the characteristic signature of a non-Hermitian exceptional point.  Thus, $D_{\rm EP}=1$ marks the transition from an entirely real spectrum to complex quasiparticle energies and is therefore the $\mathcal{RT}$-symmetry-breaking boundary.

For $D>1$, Eq.~\eqref{eq:spectrum} contains complex quasiparticle energies, so the entire $D>1$ region is $\mathcal{RT}$ broken.  The real line gap is
\begin{equation}
\Delta_{\rm Re}=\min_{k,\nu}\operatorname{Re}E_{\nu,+}(k)=2|h|.
\label{eq:real_gap}
\end{equation}
It closes at $h=h_0$.  On this line, the spectrum becomes
\begin{equation}
E^2(k)=4J^2\left[
\gamma^2\cos^2k-(D^2-1)\sin^2k\right].
\label{eq:zero_field_spectrum}
\end{equation}
The real and imaginary energies meet at
\begin{equation}
k_{\rm EP}=\pm\tan^{-1}\!\left(
\frac{|\gamma|}{\sqrt{D^2-1}}\right),
\label{eq:kEP}
\end{equation}
and the energies are purely imaginary for $|k|>|k_{\rm EP}|$. Figures~\ref{fig:spectrum_transition}(a) and~\ref{fig:spectrum_transition}(b) show the real and imaginary quasiparticle energies at $h=h_0$. The imaginary energies appear for $D>D_{\rm EP}$, and the white curves agree with the exceptional momenta in Eq.~\eqref{eq:kEP}.

The $+z$-FM and $-z$-FM phases meet at $h=h_0$.  We now determine the order of this transition from the ground-state energy.  Let $E_{\nu,+}(k)$ denote the quasiparticle energies with positive real parts.  The real part of the ground-state energy per site is
\begin{equation}
\frac{\operatorname{Re}E_0(h)}{N}=-\frac{1}{4\pi}
\operatorname{Re}\int_{-\pi/2}^{\pi/2}dk
\sum_{\nu=\pm1}E_{\nu,+}(k).
\label{eq:e0_integral}
\end{equation}
Near $h=h_0$, Eq.~\eqref{eq:e0_integral} gives
\begin{equation}
\frac{\operatorname{Re}E_0(h)}{N}
=\frac{\operatorname{Re}E_0(0)}{N}
-\frac{\sqrt{D^2-1}}{\sqrt{D^2-1+\gamma^2}}|h|
+\mathcal O(h^2).
\label{eq:e0_cusp}
\end{equation}
The $|h|$ term originates from the purely imaginary quasiparticle modes.  Positive and negative fields select different ground states, producing a cusp in $\operatorname{Re}E_0$ at $h=h_0$.  Since the field term is $-h\sum_j\sigma_j^z$, the longitudinal magnetization is the negative derivative of the ground-state energy with respect to $h$.  For the non-Hermitian Hamiltonian, the Hellmann--Feynman relation uses the right and left ground states $|R_0\rangle$ and $\langle L_0|$:
\begin{equation}
M_z=\frac{1}{N}\sum_j\operatorname{Re}
\frac{\langle L_0|\sigma_j^z|R_0\rangle}
{\langle L_0|R_0\rangle}
=-\frac{1}{N}\frac{\partial\operatorname{Re}E_0}{\partial h}.
\label{eq:mz_definition}
\end{equation}
As $h$ approaches $h_0$ from either side, the longitudinal magnetization becomes
\begin{equation}
M_z(h_0^{\pm})=
\pm\frac{\sqrt{D^2-1}}{\sqrt{D^2-1+\gamma^2}}.
\label{eq:mz_jump}
\end{equation}
The finite and opposite values of $M_z$ on the two sides of $h=h_0$ show a discontinuous reversal of the longitudinal magnetization.  We thus find two oppositely polarized $z$-FM phases in the $\mathcal{RT}$-broken region, separated by the first-order transition line $h=h_0$.  The quasiparticle spectrum further locates the $\mathcal{RT}$-symmetry-breaking line at $D=D_{\rm EP}=1$.

\section{Quantum-information probes of the phase diagram}
\label{sec:information}

Based on the phase diagram obtained above, we now examine how quantum-information quantities characterize the magnetic and non-Hermitian transitions of this model.  Since the left and right ground states of a non-Hermitian Hamiltonian differ, we use two density operators.  The self-normal (RR) density operator is constructed from the right ground state, whereas the biorthogonal (LR) density operator combines the left and right ground states,
\begin{equation}
\rho_{\rm RR}=\frac{|R_0\rangle\langle R_0|}{\langle R_0|R_0\rangle},
\qquad
\rho_{\rm LR}=\frac{|R_0\rangle\langle L_0|}{\langle L_0|R_0\rangle}.
\label{eq:RR_LR_density}
\end{equation}
Here $|R_0\rangle$ is the right ground state with the lowest real energy, and $\langle L_0|$ is the corresponding left ground state.  At $h=h_0$ for $D>1$, we take the limit $h\rightarrow0^+$.  We first study the self-normal entanglement entropy, quantum discord, and maximum quantum coherence.  We then compare them with the biorthogonal quantities to examine whether the LR definition gives a clearer characterization of the transitions governed by non-Hermiticity.  The occupied BdG states give the correlation matrix and reduced density operators in the thermodynamic limit.  The two nearest-neighbor bonds in one unit cell are inequivalent, so the QD and coherence values shown below are averaged over both bonds.

\subsection{Self-normal entanglement entropy}

\begin{figure}[!t]
	\centering
	\includegraphics[width=\columnwidth]{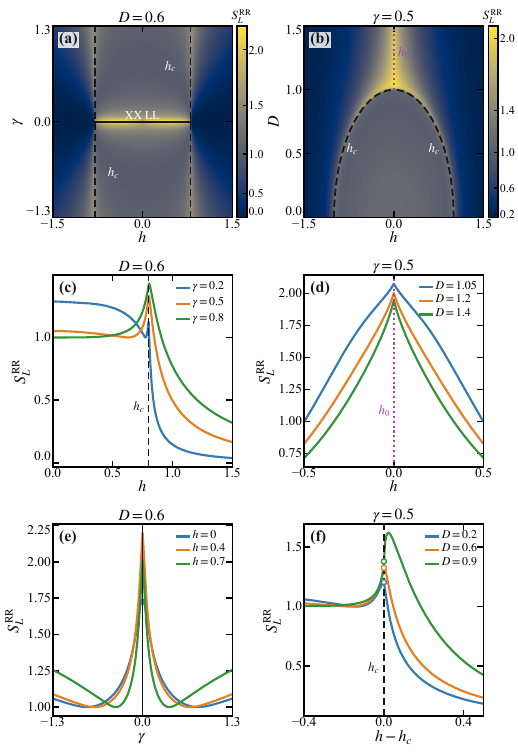}
	\caption{Self-normal block entanglement entropy $S_L^{\rm RR}$ for $L=16$ and $J=1$. (a) $(h,\gamma)$ map at $D=0.6$.  (b) $(h,D)$ map at $\gamma=0.5$. (c) Field dependence at $D=0.6$ for three anisotropies.  (d) Field dependence at $\gamma=0.5$ for three values of $D>1$.  (e) Anisotropy dependence at $D=0.6$ for three fields inside the ordered region. (f) Field dependence at $\gamma=0.5$ for three values of $D<1$.  The black dashed curves mark $h=\pm h_c$, the black solid segment marks $\gamma=0$, and the purple dotted segment marks $h=h_0$.  Open circles in (f) mark the analytic transition points.}
	\label{fig:entropy}
\end{figure}

For a block of $L$ consecutive spins, the self-normal entanglement entropy is
\begin{equation}
S_L^{\rm RR}=-\operatorname{Tr}\left(\rho_A^{\rm RR}\log_2\rho_A^{\rm RR}\right),
\qquad
\rho_A^{\rm RR}=\operatorname{Tr}_{\bar A}\rho_{\rm RR}.
\label{eq:entropy_RR}
\end{equation}
After the Jordan--Wigner transformation, the right ground state is Gaussian, so the block entropy can be obtained from the Majorana covariance matrix~\cite{Peschel2003,PeschelEisler2009}.  Using
\begin{equation}
w_{2j-1}=c_j+c_j^\dagger,\qquad
w_{2j}=-\mathrm{i}(c_j-c_j^\dagger),
\end{equation}
we define
\begin{equation}
\Gamma_{mn}^{\rm RR}=\frac{\mathrm{i}}{2}\operatorname{Tr}
\left(\rho_{\rm RR}[w_m,w_n]\right),
\qquad m,n=1,\ldots,2L.
\label{eq:majorana_covariance}
\end{equation}
If the eigenvalues of $\mathrm{i}\Gamma_A^{\rm RR}$ are $\pm\nu_\mu$, with $0\leq\nu_\mu\leq1$, the entropy is
\begin{align}
S_L^{\rm RR}=-\sum_{\mu=1}^{L}\sum_{s=\pm}
\frac{1+s\nu_\mu}{2}
\log_2\!\left(\frac{1+s\nu_\mu}{2}\right).
\label{eq:entropy_covariance}
\end{align}
We use $L=16$ in the phase maps and one-dimensional cuts.

For $D<1$, Fig.~\ref{fig:entropy}(a) displays the magnetic boundaries $h=\pm h_c$ and the anisotropy transition at $\gamma=0$.  The field cuts in Fig.~\ref{fig:entropy}(c) change sharply at $h_c=0.8$, and the anisotropy cuts in Fig.~\ref{fig:entropy}(e) peak at $\gamma=0$.  As $D$ increases, the transition points in Fig.~\ref{fig:entropy}(f) follow the analytic magnetic boundary in Eq.~\eqref{eq:hc}.  Thus, $S_L^{\rm RR}$ identifies the magnetic transitions for $D<1$.

For $D>1$, Fig.~\ref{fig:entropy}(b) displays the first-order boundary $h=h_0$.  All field cuts in Fig.~\ref{fig:entropy}(d) develop a cusp at $h=h_0$, which follows from the abrupt reversal of the longitudinal magnetization.  The self-normal block entropy therefore detects the first-order transition between the $+z$-FM and $-z$-FM phases.  However, $S_L^{\rm RR}$ remains smooth at $D=D_{\rm EP}=1$ and therefore fails to identify the $\mathcal{RT}$-symmetry-breaking transition or the accompanying PM--$z$-FM transition.

\subsection{Self-normal quantum discord}

The two-site mutual information is
\begin{equation}
I(\rho_{AB}^{\rm RR})=S(\rho_A^{\rm RR})+S(\rho_B^{\rm RR})
-S(\rho_{AB}^{\rm RR}).
\label{eq:mutual_information}
\end{equation}
After a projective measurement $\{\Pi_A^\mu\}$ on spin $A$, the outcome probability and the conditional state of spin $B$ are
\begin{align}
p_\mu={}&\operatorname{Tr}
\left[(\Pi_A^\mu\otimes\mathbb I)\rho_{AB}^{\rm RR}\right],\nonumber\\
\rho_{B|\mu}^{\rm RR}={}&\frac{1}{p_\mu}\operatorname{Tr}_A\Big[
(\Pi_A^\mu\otimes\mathbb I)\rho_{AB}^{\rm RR}
(\Pi_A^\mu\otimes\mathbb I)\Big].
\label{eq:conditional_state}
\end{align}
The self-normal quantum discord is~\cite{Ollivier2001,Henderson2001}
\begin{equation}
\mathrm{QD}^{\rm RR}=S(\rho_A^{\rm RR})-S(\rho_{AB}^{\rm RR})
+\min_{\{\Pi_A^\mu\}}\sum_\mu p_\mu S(\rho_{B|\mu}^{\rm RR}).
\label{eq:QD}
\end{equation}

\begin{figure}[!t]
\centering
\includegraphics[width=\columnwidth]{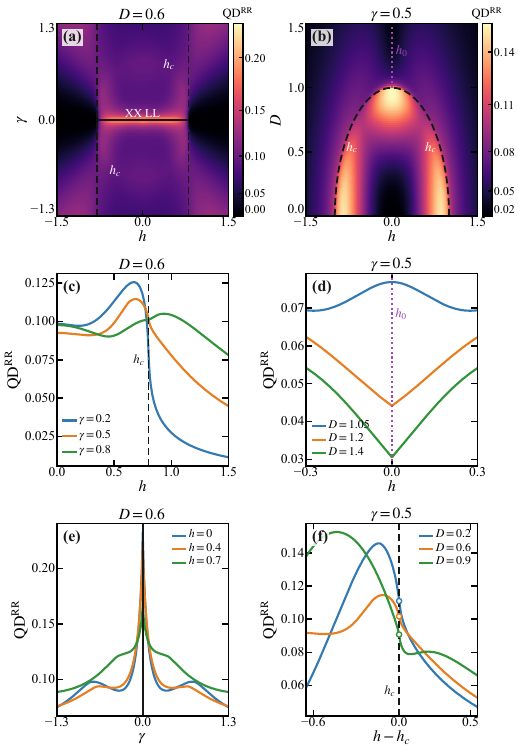}
\caption{Self-normal nearest-neighbor quantum discord $\mathrm{QD}^{\rm RR}$ at $J=1$, averaged over the two alternating bonds.  The parameter planes and cuts are the same as in Fig.~\ref{fig:entropy}.  The black dashed curves mark $h=\pm h_c$, the black solid segment marks $\gamma=0$, and the purple dotted segment marks $h=h_0$.  Open circles in (f) mark the analytic transition points.}
\label{fig:discord}
\end{figure}

The self-normal QD gives the same transition signals as the block entropy.  For $D<1$, Fig.~\ref{fig:discord}(a) displays the magnetic boundaries $h=\pm h_c$ and the anisotropy transition at $\gamma=0$.  The field cuts in Fig.~\ref{fig:discord}(c) change slope at $h_c=0.8$, and the marked transition points in Fig.~\ref{fig:discord}(f) follow the analytic magnetic boundary in Eq.~\eqref{eq:hc} as $D$ varies.  The anisotropy cuts in Fig.~\ref{fig:discord}(e) peak at $\gamma=0$.  Thus, $\mathrm{QD}^{\rm RR}$ identifies both the $x$-FM--$y$-FM transition and the ordered-to-paramagnetic transition.

For $D>1$, Fig.~\ref{fig:discord}(b) displays the first-order boundary $h=h_0$.  All field cuts in Fig.~\ref{fig:discord}(d) develop a cusp at $h=h_0$.  These cusps coincide with the energy cusp and magnetization jump in Fig.~\ref{fig:spectrum_transition}.  Thus, $\mathrm{QD}^{\rm RR}$ detects the first-order transition between the $+z$-FM and $-z$-FM phases.  However, like $S_L^{\rm RR}$, $\mathrm{QD}^{\rm RR}$ remains smooth at $D=D_{\rm EP}=1$ and therefore fails to identify the $\mathcal{RT}$-symmetry-breaking transition or the accompanying PM--$z$-FM transition.

\subsection{Self-normal maximum quantum coherence}
\label{sec:qcmax}

We use the observable-based coherence introduced in Ref.~\cite{Girolami2014}.  For a unit vector $\bm n$, the quantum coherence associated with the local spin direction $\bm n\!\cdot\!\bm\sigma$ is
\begin{equation}
\mathrm{QC}(\rho,\bm n)=-\frac14\operatorname{Tr}
\left[\rho,\bm n\!\cdot\!\bm\sigma\otimes\mathbb I\right]^2.
\label{eq:Q_direction}
\end{equation}
For a fixed direction $\bm n$, $\mathrm{QC}(\rho,\bm n)$ measures the noncommutativity between the density operator and the corresponding local spin component.  It vanishes when the two operators commute.

Previous work introduced the maximum quantum coherence by maximizing $\mathrm{QC}(\rho,\bm n)$ over all local spin directions and calculated this maximum by scanning the direction angles~\cite{Lv2022}.  For a given reduced density matrix, a direct scan requires calculating the coherence point by point over the entire unit sphere. However, a scan with a finite angular step may miss the true optimal direction, while further reducing the step increases the computational cost.

We convert this maximization over directions into a Rayleigh-quotient problem by constructing a $3\times3$ coherence matrix.  First, expanding the commutator along the $x$, $y$, and $z$ directions gives
\begin{equation}
\left[\rho,\bm n\!\cdot\!\bm\sigma\otimes\mathbb I\right]
=\sum_{\alpha=x,y,z}n_\alpha
\left[\rho,\sigma_\alpha\otimes\mathbb I\right].
\label{eq:coherence_commutator_expansion}
\end{equation}
Substituting this expansion into the definition of coherence gives
\begin{equation}
\begin{aligned}
\mathrm{QC}(\rho,\bm n)
={}&-\frac14\sum_{\alpha,\beta=x,y,z}n_\alpha n_\beta\\
&\times\operatorname{Tr}\left(
[\rho,\sigma_\alpha\otimes\mathbb I]
[\rho,\sigma_\beta\otimes\mathbb I]\right).
\end{aligned}
\label{eq:coherence_expanded}
\end{equation}
We write the coefficients independent of $\bm n$ as the elements of the coherence matrix,
\begin{equation}
M_{\alpha\beta}=-\frac14\operatorname{Tr}
\left([\rho,\sigma_\alpha\otimes\mathbb I]
[\rho,\sigma_\beta\otimes\mathbb I]\right),
\qquad \alpha,\beta=x,y,z.
\label{eq:coherence_matrix}
\end{equation}
The coherence along an arbitrary direction can then be written as
\begin{equation}
\mathrm{QC}(\rho,\bm n)=
\sum_{\alpha,\beta=x,y,z}n_\alpha M_{\alpha\beta}n_\beta
=\bm n^{\mathsf T}M\bm n.
\label{eq:coherence_quadratic}
\end{equation}
For the self-normal density operator $\rho_{\rm RR}$, $M^{\rm RR}$ is real and symmetric.  By the Rayleigh-quotient theorem,
\begin{equation}
\begin{aligned}
\mathrm{QC}_{\max}^{\rm RR}
&=\max_{\bm n^{\mathsf T}\bm n=1}\mathrm{QC}(\rho_{\rm RR},\bm n)\\
&=\max_{\substack{\bm n\in\mathbb R^3\\\bm n\ne0}}
\frac{\bm n^{\mathsf T}M^{\rm RR}\bm n}{\bm n^{\mathsf T}\bm n}\\
&=\lambda_{\max}(M^{\rm RR}).
\end{aligned}
\label{eq:QCmax}
\end{equation}
The corresponding normalized eigenvector gives the optimal spin direction.  Therefore, for a given reduced density matrix, we only need to construct $M^{\rm RR}$, which contains nine matrix elements, and solve a $3\times3$ eigenvalue problem to obtain the maximum coherence and its optimal direction. Furthermore, since $M^{\rm RR}$ is real and symmetric, only six matrix elements are independent, so calculating these six elements is sufficient to carry out the above maximization, greatly reducing the computational cost. Since no scan over the direction angles is required, this method also avoids missing the true optimal direction and maximum due to angular discretization.

\begin{figure}[!t]
\centering
\includegraphics[width=\columnwidth]{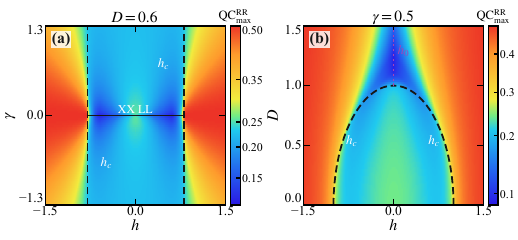}
\caption{Self-normal maximum quantum coherence $\mathrm{QC}_{\max}^{\rm RR}$ at $J=1$, averaged over the two alternating bonds.  (a) $(h,\gamma)$ map at $D=0.6$.  (b) $(h,D)$ map at $\gamma=0.5$.  The black dashed curves mark $h=\pm h_c$, the black solid segment marks $\gamma=0$, and the purple dotted segment marks $h=h_0$.}
\label{fig:qcmax}
\end{figure}

The self-normal maximum quantum coherence gives the same transition signals as the block entropy and QD\@.  For $D<1$, Fig.~\ref{fig:qcmax}(a) displays the magnetic boundaries $h=\pm h_c$ and the anisotropy transition at $\gamma=0$.  Thus, $\mathrm{QC}_{\max}^{\rm RR}$ identifies both the $x$-FM--$y$-FM transition and the ordered-to-paramagnetic transition.  For $D>1$, Fig.~\ref{fig:qcmax}(b) displays the first-order boundary $h=h_0$.  Thus, $\mathrm{QC}_{\max}^{\rm RR}$ detects the first-order transition between the $+z$-FM and $-z$-FM phases.  However, like $S_L^{\rm RR}$ and $\mathrm{QD}^{\rm RR}$, $\mathrm{QC}_{\max}^{\rm RR}$ remains smooth at $D=D_{\rm EP}=1$.

Our analysis suggests that the common behavior of the three self-normal quantum-information quantities at $D=D_{\rm EP}$ follows from the right-state projector.  Let $R_k=(|r_{k,+}\rangle,|r_{k,-}\rangle)$ contain the two right BdG modes that determine the ground-state correlation matrix.  For $D\neq1$, their contribution to the self-normal correlation matrix is
\begin{equation}
P_{\rm RR}(k)=
R_k\left(R_k^\dagger R_k\right)^{-1}R_k^\dagger.
\label{eq:RR_projector}
\end{equation}
Near $D=1$, the two coalescing right eigenstates have the form
\begin{equation}
|r_{k,\pm}\rangle=
|r_{k,0}\rangle
\pm\sqrt{D-1}\,|r_{k,1}\rangle
+\mathcal O(D-1).
\label{eq:RR_EP_modes}
\end{equation}
Although each eigenstate contains a term proportional to $\sqrt{D-1}$, both states enter $P_{\rm RR}(k)$.  Their sum and their difference divided by $2\sqrt{D-1}$ form a regular basis of the same subspace.  Since the orthogonal projector is unchanged by this basis transformation, the square-root terms cancel in $P_{\rm RR}(k)$, giving
\begin{equation}
P_{\rm RR}(k,D)=
P_{\rm RR}(k,1)+\mathcal O(D-1).
\label{eq:RR_projector_smooth}
\end{equation}
The self-normal reduced density operators therefore remain smooth at $D=1$.  In addition, $M^{\rm RR}$ is real and symmetric, so its eigenvalues do not undergo an exceptional complex splitting.  The left--right eigenstate overlap that vanishes at the exceptional point does not enter $\rho_{\rm RR}$.  The self-normal entanglement entropy, QD, and maximum quantum coherence therefore fail to identify the $\mathcal{RT}$-symmetry-breaking transition and the accompanying PM--$z$-FM transition at $D=D_{\rm EP}$.

Based on the above analysis, because the biorthogonal definition contains both the left and right eigenstates and retains the non-Hermitian information carried by their overlap, it may be able to identify the $\mathcal{RT}$-symmetry-breaking transition missed by the self-normal quantum-information quantities.  In the next subsection, we therefore calculate the biorthogonal quantum-information quantities and determine which of them identifies this exceptional boundary.

\subsection{Biorthogonal quantum-information quantities}
\label{sec:LR_information}

In light of the above possible reasons for the failure to detect the exceptional point, we turn to biorthogonal quantum-information quantities and first examine the biorthogonal block entropy.  It is defined as
\begin{equation}
S_L^{\rm LR}=-\operatorname{Tr}\left(\rho_A^{\rm LR}\log_2\rho_A^{\rm LR}\right),
\qquad
\rho_A^{\rm LR}=\operatorname{Tr}_{\bar A}\rho_{\rm LR}.
\label{eq:entropy_LR}
\end{equation}

\begin{figure}[!t]
\centering
\includegraphics[width=\columnwidth]{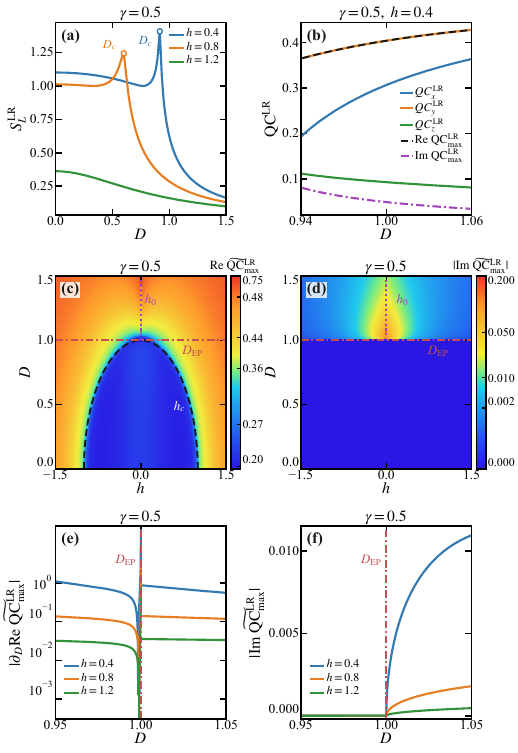}
\caption{Biorthogonal entropy and coherence at $\gamma=0.5$ and $J=1$.  (a) $S_L^{\rm LR}$ as a function of $D$ for three fields.  The open circles mark $D_c=\sqrt{1-h^2}$ for $h=0.4$ and $0.8$.  (b) $QC_x^{\rm LR}$, $QC_y^{\rm LR}$, $QC_z^{\rm LR}$, $\operatorname{Re}\mathrm{QC}_{\max}^{\rm LR}$, and $\operatorname{Im}\mathrm{QC}_{\max}^{\rm LR}$ near $D=D_{\rm EP}$ at $h=0.4$.  (c) $\operatorname{Re}\widetilde{\mathrm{QC}}_{\max}^{\rm LR}$ and (d) $|\operatorname{Im}\widetilde{\mathrm{QC}}_{\max}^{\rm LR}|$ in the $(h,D)$ plane.  (e) $|\partial_D\operatorname{Re}\widetilde{\mathrm{QC}}_{\max}^{\rm LR}|$ on a logarithmic scale and (f) $|\operatorname{Im}\widetilde{\mathrm{QC}}_{\max}^{\rm LR}|$ across $D=D_{\rm EP}$ for three fields.  The black dashed curves in (c) mark $h=\pm h_c$, the purple dotted segments in (c) and (d) mark $h=h_0$, and the red dash-dotted lines in (c)--(f) mark $D=D_{\rm EP}$.  The coherence is averaged over the two alternating bonds.}
\label{fig:LR_information}
\end{figure}

Figure~\ref{fig:LR_information}(a) shows $S_L^{\rm LR}$ as a function of $D$ for three transverse fields.  For $h=0.4$ and $0.8$, the sharp peaks occur at the magnetic boundaries $D_c=\sqrt{1-h^2}$.  The biorthogonal entropy therefore identifies the ferromagnetic-to-paramagnetic transition obtained from the exact mapping.  But it remains real and smooth at $D=D_{\rm EP}$ and fails to identify the exceptional boundary.

The reason is that the eigenvalues of the reduced correlation matrix occur in pairs.  Near $D=1$, the two eigenvalues that merge can be written as
\begin{equation}
\xi_\pm=\xi_0\pm\delta\xi.
\label{eq:LR_entropy_pair}
\end{equation}
For $D>1$, $\delta\xi$ becomes imaginary, so that $\xi_+$ and $\xi_-$ form a complex-conjugate pair.  The entropy contains both eigenvalues through $s(\xi)=-\xi\log_2\xi-(1-\xi)\log_2(1-\xi)$.  Expanding their sum around $\xi_0$ gives
\begin{equation}
s(\xi_+)+s(\xi_-)=2s(\xi_0)+\mathcal O\!\left[(\delta\xi)^2\right].
\label{eq:LR_entropy_cancellation}
\end{equation}
The terms linear in $\delta\xi$ cancel, so the leading change of the eigenvalues near the exceptional boundary does not enter the block entropy.  Although $S_L^{\rm LR}$ is defined using both the left and right eigenstates, it still fails to detect the exceptional transition.

We next examine whether biorthogonal quantum coherence can identify the exceptional boundary.  Equation~\eqref{eq:coherence_quadratic} expresses the coherence along an arbitrary spin direction $\bm n$ as a quadratic form.  Replacing $\rho_{\rm RR}$ with $\rho_{\rm LR}$ gives the complex symmetric coherence matrix $M^{\rm LR}$ for the biorthogonal state,
\begin{equation}
\mathrm{QC}^{\rm LR}(\bm n)=\bm n^{\mathsf T}M^{\rm LR}\bm n.
\label{eq:QC_general_LR}
\end{equation}
To show separately how the real and imaginary parts of the coherence arise, we write
\begin{equation}
M^{\rm LR}=M_H^{\rm LR}+\mathrm{i} M_I^{\rm LR},
\label{eq:coherence_LR_decomposition}
\end{equation}
where
\begin{equation}
M_H^{\rm LR}=\frac{M^{\rm LR}+M^{{\rm LR}\dagger}}{2},
\qquad
M_I^{\rm LR}=\frac{M^{\rm LR}-M^{{\rm LR}\dagger}}{2\mathrm{i}}.
\label{eq:coherence_LR_HI}
\end{equation}

Previous work on the $XXZ$ chain with a staggered imaginary field showed that the biorthogonal coherence $QC_x^{\rm LR}$ along the $x$ direction can itself acquire an imaginary part and identify the $\mathcal{PT}$-symmetry-breaking transition~\cite{ZhangYu2025}.  The coherence-matrix decomposition above clearly explains the origin of this result: in that model, the staggered imaginary field produces complex left--right spin correlations, and the resulting non-Hermitian contribution appears directly in a diagonal element of the biorthogonal coherence matrix.  Consequently, the $xx$ diagonal element $M_{I,xx}^{\rm LR}$ of $M_I^{\rm LR}$ is nonzero, so $QC_x^{\rm LR}=M_{xx}^{\rm LR}=M_{H,xx}^{\rm LR}+\mathrm{i} M_{I,xx}^{\rm LR}$ directly contains the non-Hermitian information.

In our model, however, the staggered imaginary DM interaction gives a different coherence-matrix structure,
\begin{equation}
\begin{aligned}
M_H^{\rm LR}&=
\begin{pmatrix}
QC_x^{\rm LR}&0&0\\
0&QC_y^{\rm LR}&0\\
0&0&QC_z^{\rm LR}
\end{pmatrix},\\[3pt]
M_I^{\rm LR}&=
\begin{pmatrix}
0&M_{I,xy}^{\rm LR}&0\\
M_{I,xy}^{\rm LR}&0&0\\
0&0&0
\end{pmatrix}.
\end{aligned}
\label{eq:coherence_LR_structure}
\end{equation}
The three Cartesian coherences lie on the diagonal of $M_H^{\rm LR}$, while the imaginary coherence appears only in the $xy$ off-diagonal element of $M_I^{\rm LR}$.  The biorthogonal coherences along the fixed Cartesian directions are therefore
\begin{equation}
QC_\alpha^{\rm LR}=M_{\alpha\alpha}^{\rm LR}
=M_{H,\alpha\alpha}^{\rm LR},
\qquad \alpha=x,y,z.
\label{eq:QC_cartesian_LR}
\end{equation}
Figure~\ref{fig:LR_information}(b) shows that $QC_x^{\rm LR}$, $QC_y^{\rm LR}$, and $QC_z^{\rm LR}$ all vary smoothly through $D=D_{\rm EP}$.  A coherence along a Cartesian axis reads only one diagonal element of $M_H^{\rm LR}$, while $M_{I,xy}^{\rm LR}$, which carries the imaginary coherence, does not enter any of these three quantities.  The biorthogonal coherences along fixed Cartesian directions therefore still fail to identify the $\mathcal{RT}$-symmetry-breaking transition.

Since $M^{\rm LR}$ is complex symmetric, both $M_H^{\rm LR}$ and $M_I^{\rm LR}$ are real symmetric matrices.  For a real unit vector $\bm n$, the real and imaginary parts of the biorthogonal coherence are
\begin{align}
\operatorname{Re}\mathrm{QC}^{\rm LR}(\bm n)
&=\bm n^{\mathsf T}M_H^{\rm LR}\bm n,\nonumber\\
\operatorname{Im}\mathrm{QC}^{\rm LR}(\bm n)
&=\bm n^{\mathsf T}M_I^{\rm LR}\bm n.
\label{eq:coherence_LR_parts}
\end{align}
The Rayleigh-quotient method used for $M^{\rm RR}$ in Sec.~\ref{sec:qcmax} can therefore still be applied separately to $M_H^{\rm LR}$ and $M_I^{\rm LR}$.  We maximize the real and imaginary parts of the coherence separately over all real spin directions and denote their maxima by
\begin{align}
\operatorname{Re}\mathrm{QC}_{\max}^{\rm LR}
&=\lambda_{\max}\!\left(M_H^{\rm LR}\right),\nonumber\\
\operatorname{Im}\mathrm{QC}_{\max}^{\rm LR}
&=\lambda_{\max}\!\left(M_I^{\rm LR}\right).
\label{eq:coherence_LR_extrema}
\end{align}
For the parameters used in Fig.~\ref{fig:LR_information}(b),
\begin{equation}
\operatorname{Re}\mathrm{QC}_{\max}^{\rm LR}
=\lambda_{\max}\!\left(M_H^{\rm LR}\right)
=QC_y^{\rm LR},
\label{eq:Re_QCmax_LR}
\end{equation}
and the largest real coherence lies along the $y$ direction.  On the other hand,
\begin{equation}
\operatorname{Im}\mathrm{QC}^{\rm LR}(\bm n)
=2n_xn_yM_{I,xy}^{\rm LR},
\label{eq:Im_Q_LR_direction}
\end{equation}
which gives
\begin{equation}
\operatorname{Im}\mathrm{QC}_{\max}^{\rm LR}
=\lambda_{\max}\!\left(M_I^{\rm LR}\right)
=\left|M_{I,xy}^{\rm LR}\right|.
\label{eq:Im_QCmax_LR}
\end{equation}
The directions
\begin{equation}
\bm n=\frac{\hat{\bm x}\pm\hat{\bm y}}{\sqrt{2}}
\label{eq:Im_QCmax_directions}
\end{equation}
give the two extrema $\pm\left|M_{I,xy}^{\rm LR}\right|$.  The largest real and imaginary coherences therefore occur along different spin directions.  However, Fig.~\ref{fig:LR_information}(b) shows that both vary smoothly through $D=1$.  Separately maximizing the real and imaginary coherences over real spin directions still fails to identify the exceptional boundary.

To retain the real $x$--$y$ coherence difference and the imaginary $xy$ coherence in the same quantity, and to examine their competition, we further study the eigenvalues of the full complex coherence matrix $M^{\rm LR}$.  For this purpose, we retain the original transpose product and extend $\bm n$ from a real vector to a complex vector, using the bilinear Rayleigh quotient for complex symmetric matrices~\cite{ArbenzHochstenbach2004}:
\begin{equation}
\frac{\bm n^{\mathsf T}M^{\rm LR}\bm n}{\bm n^{\mathsf T}\bm n},
\qquad \bm n\in\mathbb C^3,\quad
\bm n^{\mathsf T}\bm n\ne0.
\label{eq:coherence_LR_Rayleigh}
\end{equation}
Since $M^{\rm LR}$ is complex symmetric, the stationarity condition for this quotient is
\begin{equation}
M^{\rm LR}\bm n
=\frac{\bm n^{\mathsf T}M^{\rm LR}\bm n}{\bm n^{\mathsf T}\bm n}\,\bm n.
\label{eq:coherence_LR_stationary}
\end{equation}
Therefore, the stationary values of this quotient correspond to the eigenvalues of the full complex coherence matrix.  For the present model, $M^{\rm LR}$ takes the form
\begin{equation}
M^{\rm LR}=
\begin{pmatrix}
QC_x^{\rm LR}&\mathrm{i} M_{I,xy}^{\rm LR}&0\\
\mathrm{i} M_{I,xy}^{\rm LR}&QC_y^{\rm LR}&0\\
0&0&QC_z^{\rm LR}
\end{pmatrix}.
\label{eq:coherence_matrix_full_LR}
\end{equation}
Its three coherence eigenvalues are
\begin{equation}
\begin{aligned}
\mathrm{QC}_{\lambda,\pm}^{\rm LR}
={}&\frac{QC_x^{\rm LR}+QC_y^{\rm LR}}{2}\\
&\pm\sqrt{
\left(\frac{QC_x^{\rm LR}-QC_y^{\rm LR}}{2}\right)^2
-\left(M_{I,xy}^{\rm LR}\right)^2},\\[3pt]
\mathrm{QC}_{\lambda,z}^{\rm LR}
={}&QC_z^{\rm LR}.
\end{aligned}
\label{eq:QC_LR_eigenvalues}
\end{equation}
In the following, we use $\widetilde{\mathrm{QC}}_{\max}^{\rm LR}$ to denote the eigenvalue with the largest real part among the three eigenvalues of $M^{\rm LR}$ and show it in Fig.~\ref{fig:LR_information}.  If several eigenvalues share the largest real part, we choose the one with the largest imaginary part.

We find that the first term inside the square root comes from the difference between the real $x$- and $y$-direction coherences in $M_H^{\rm LR}$, whereas the second comes from the imaginary $xy$ coherence in $M_I^{\rm LR}$.  The square root as a whole therefore reflects the competition between the real coherence difference along the $x$ and $y$ directions and the imaginary coherence connecting these two directions.  Our calculations show that the two contributions become exactly balanced at $D=D_{\rm EP}=1$, where
\begin{equation}
\frac{\left|QC_x^{\rm LR}-QC_y^{\rm LR}\right|}{2}
=\left|M_{I,xy}^{\rm LR}\right|.
\label{eq:QC_LR_balance}
\end{equation}
The eigenvalues of the full complex coherence matrix therefore contain the information from both $M_H^{\rm LR}$ and $M_I^{\rm LR}$.  In this model, their balance occurs exactly at the exceptional boundary obtained from the quasiparticle spectrum and therefore identifies the location of the $\mathcal{RT}$-symmetry-breaking transition.  Their competition also directly determines whether the two coherence eigenvalues are real or complex.  For $D<1$, the difference between the real $x$- and $y$-direction coherences is larger than the imaginary $xy$ coherence, and both eigenvalues are real.  At $D=D_{\rm EP}$, the two contributions become equal, the square root vanishes, and the two coherence eigenvalues and their eigenvectors coalesce.  For $D>1$, the imaginary $xy$ coherence becomes larger, the square root becomes imaginary, and the two eigenvalues form a complex-conjugate pair.

Figures~\ref{fig:LR_information}(c) and~\ref{fig:LR_information}(d) show the real and imaginary parts of $\widetilde{\mathrm{QC}}_{\max}^{\rm LR}$ in the $(h,D)$ plane.  For $D<1$, $\widetilde{\mathrm{QC}}_{\max}^{\rm LR}$ is real.  Its $(h,\gamma)$ phase diagram agrees with the self-normal maximum quantum coherence in Fig.~\ref{fig:qcmax}(a), and both identify the same $x$-FM, $y$-FM, XX Luttinger-liquid, and paramagnetic regions.  Its real part in Fig.~\ref{fig:LR_information}(c) also clearly displays the magnetic boundaries $h=\pm h_c$.  In the $D>1$ region, it identifies the first-order boundary $h=h_0$ between the two $z$-FM phases.

The main change occurs at the exceptional boundary.  Figure~\ref{fig:LR_information}(e) shows that the first derivative of $\operatorname{Re}\widetilde{\mathrm{QC}}_{\max}^{\rm LR}$ develops a sharp peak at $D=D_{\rm EP}$.  At the same position, $\operatorname{Im}\widetilde{\mathrm{QC}}_{\max}^{\rm LR}$ becomes nonzero in Figs.~\ref{fig:LR_information}(d) and~\ref{fig:LR_information}(f).  Both signals follow from the same change: $\mathrm{QC}_{\lambda,+}^{\rm LR}$ and $\mathrm{QC}_{\lambda,-}^{\rm LR}$ merge at $D=1$ and form a complex-conjugate pair for $D>1$.  The coherence eigenvalue $\widetilde{\mathrm{QC}}_{\max}^{\rm LR}$ therefore identifies the $\mathcal{RT}$-symmetry-breaking transition and the accompanying PM--$z$-FM transition.

The above results show that $\widetilde{\mathrm{QC}}_{\max}^{\rm LR}$, which contains the non-Hermitian information carried by the left and right eigenstates and the coherence between different spin directions, gives the most complete characterization of the magnetic and $\mathcal{RT}$ phase diagrams among the quantum-information quantities studied here.

\section{Conclusion}
\label{sec:conclusion}

We investigate the $XY$ chain with a transverse field and a staggered imaginary DM interaction and determine its magnetic and $\mathcal{RT}$ phase diagrams.  For $D<1$, a staggered nonunitary rotation absorbs the imaginary DM interaction into the effective exchange and anisotropy and maps the model exactly onto the Hermitian $XY$ chain.  From this mapping, we obtain the $x$-FM, $y$-FM, XX Luttinger-liquid, and paramagnetic phases and the magnetic boundary $h_c=J\sqrt{1-D^2}$.  For $D>1$, an additional odd-site $\pi$ rotation gives a non-Hermitian $XY$ chain with imaginary anisotropy and a staggered transverse field.  At $D=D_{\rm EP}=1$, the quasiparticle eigenvalues and eigenvectors coalesce pairwise at two second-order exceptional points, and the entire $D>1$ region lies in the $\mathcal{RT}$-broken phase.  We find two oppositely polarized $z$-FM phases in this region: $+z$-FM for $h>0$ and $-z$-FM for $h<0$.  They meet at $h=h_0=0$, where the $|h|$ term in $\operatorname{Re}E_0$ produces an energy cusp and a finite jump in $M_z$, establishing a first-order transition inside the $\mathcal{RT}$-broken phase.

We further systematically characterize the performance of quantum information probes for detecting phase transitions in the studied system within self-normalized and biorthogonal theoretical frameworks. By employing the Rayleigh quotient method, the maximization of spin-orientation-resolved quantum coherence is recast as an eigenvalue problem of a $3\times3$ coherence matrix. This procedure eliminates the necessity for exhaustive angular scanning, permitting direct evaluation of the maximum quantum coherence and its optimal spin orientation and substantially lowering computational cost. Numerical results indicate that conventional observables, including entanglement entropy, quantum discord, and maximal quantum coherence obtained from the two formalisms, effectively capture the anisotropic transition at \(\gamma=0\), the magnetic boundary \(h_c\), and the first-order phase boundary \(h=h_0\). In contrast, none of these quantities are capable of detecting the \(\mathcal{RT}\)-symmetry-breaking phase transition at \(D=D_{\rm EP}\).

To remedy this fundamental limitation, this study extends the Rayleigh quotient formalism to construct a mathematically consistent maximal quantum coherence measure \(\widetilde{\mathrm{QC}}_{\max}^{\rm LR}\). This newly defined quantity integrates the complete non-Hermitian information carried by left and right eigenstates, which enables it to quantify the competition between spin-coherence contributions from the real and imaginary components of the eigenvalue equation. Benefiting from this unique property, \(\widetilde{\mathrm{QC}}_{\max}^{\rm LR}\) can effectively characterize conventional phase transitions and precisely capture the elusive \(\mathcal{RT}\) phase boundary. Nevertheless, since \(\widetilde{\mathrm{QC}}_{\max}^{\rm LR}\) is derived from a purely mathematical generalization of the Rayleigh quotient for complex-symmetric matrices, its intrinsic physical mechanism and implications remain unclear, which deserves further exploration.

\section*{Acknowledgments}
We acknowledge financial support from the National Natural Science Foundation of China (Grant Nos. 12074376 and 12505017), the Beijing National Laboratory for Condensed Matter Physics (2025BNLCMPKF017), the Beijing Municipal Natural Science Foundation (Grant No. 1222027), and the robotic AI-Scientist platform of the Chinese Academy of Sciences.

\appendix

\section{Energy cusp at \texorpdfstring{$h=h_0$}{h=h0}}
\label{app:cusp}

For $D>1$, the zero-field quasiparticle energy is purely imaginary for $|k|>|k_{\rm EP}|$.  Expanding the solutions with positive real part at small $|h|$ gives
\begin{equation}
\operatorname{Re}E_{\nu,+}(k)=
\frac{2\sqrt{D^2-1}|\sin k|}
{\sqrt{(D^2-1)\sin^2k-\gamma^2\cos^2k}}
|h|+\mathcal O(h^2).
\label{eq:small_h_energy}
\end{equation}
The required momentum integral is
\begin{align}
&\int_{|k|>|k_{\rm EP}|}
\frac{\sqrt{D^2-1}|\sin k|}
{\sqrt{(D^2-1)\sin^2k-\gamma^2\cos^2k}}\,dk
\nonumber\\
&\hspace{2.3cm}=\frac{\pi\sqrt{D^2-1}}
{\sqrt{D^2-1+\gamma^2}}.
\label{eq:cusp_integral}
\end{align}
Substitution into Eq.~\eqref{eq:e0_integral} gives Eq.~\eqref{eq:e0_cusp}.  Differentiation with respect to $h$ gives the one-sided longitudinal magnetization in Eq.~\eqref{eq:mz_jump}.

\bibliography{references}

@article{JordanWigner1928,
  author = {Jordan, P. and Wigner, E.},
  title = {{{\"U}ber das Paulische {\"A}quivalenzverbot}},
  journal = {Z. Phys.},
  volume = {47},
  pages = {631--651},
  year = {1928},
  doi = {10.1007/BF01331938}
}

@article{Lieb1961,
  author = {Lieb, Elliott and Schultz, Theodore and Mattis, Daniel},
  title = {Two Soluble Models of an Antiferromagnetic Chain},
  journal = {Ann. Phys. (N.Y.)},
  volume = {16},
  pages = {407--466},
  year = {1961},
  doi = {10.1016/0003-4916(61)90115-4}
}

@article{BarouchMcCoyDresden1970,
  author = {Barouch, Eytan and McCoy, Barry M. and Dresden, Max},
  title = {Statistical Mechanics of the {XY} Model. I},
  journal = {Phys. Rev. A},
  volume = {2},
  pages = {1075--1092},
  year = {1970},
  doi = {10.1103/PhysRevA.2.1075}
}

@article{Barouch1971,
  author = {Barouch, Eytan and McCoy, Barry M.},
  title = {Statistical Mechanics of the {XY} Model. {II}. Spin-Correlation Functions},
  journal = {Phys. Rev. A},
  volume = {3},
  pages = {786--804},
  year = {1971},
  doi = {10.1103/PhysRevA.3.786}
}

@article{Ashida2020,
  author = {Ashida, Yuto and Gong, Zongping and Ueda, Masahito},
  title = {Non-{H}ermitian Physics},
  journal = {Adv. Phys.},
  volume = {69},
  pages = {249--435},
  year = {2020},
  doi = {10.1080/00018732.2021.1876991}
}

@article{DingFangMa2022,
  author = {Ding, Kun and Fang, Chen and Ma, Guancong},
  title = {Non-{H}ermitian Topology and Exceptional-Point Geometries},
  journal = {Nat. Rev. Phys.},
  volume = {4},
  pages = {745--760},
  year = {2022},
  doi = {10.1038/s42254-022-00516-5}
}

@article{MedenGrunwaldKennes2023,
  author = {Meden, Volker and Grunwald, Lisa and Kennes, Dante M.},
  title = {{PT}-Symmetric, Non-{H}ermitian Quantum Many-Body Physics---A Methodological Perspective},
  journal = {Rep. Prog. Phys.},
  volume = {86},
  pages = {124501},
  year = {2023},
  doi = {10.1088/1361-6633/ad05f3}
}

@article{Bender1998,
  author = {Bender, Carl M. and Boettcher, Stefan},
  title = {Real Spectra in Non-{H}ermitian Hamiltonians Having {$\mathcal{PT}$} Symmetry},
  journal = {Phys. Rev. Lett.},
  volume = {80},
  pages = {5243--5246},
  year = {1998},
  doi = {10.1103/PhysRevLett.80.5243}
}

@article{Heiss2012,
  author = {Heiss, W. D.},
  title = {The Physics of Exceptional Points},
  journal = {J. Phys. A: Math. Theor.},
  volume = {45},
  pages = {444016},
  year = {2012},
  doi = {10.1088/1751-8113/45/44/444016}
}

@article{Bergholtz2021,
  author = {Bergholtz, Emil J. and Budich, Jan Carl and Kunst, Flore K.},
  title = {Exceptional Topology of Non-{H}ermitian Systems},
  journal = {Rev. Mod. Phys.},
  volume = {93},
  pages = {015005},
  year = {2021},
  doi = {10.1103/RevModPhys.93.015005}
}

@article{Wiersig2023,
  author = {Wiersig, Jan},
  title = {Petermann Factors and Phase Rigidities near Exceptional Points},
  journal = {Phys. Rev. Research},
  volume = {5},
  pages = {033042},
  year = {2023},
  doi = {10.1103/PhysRevResearch.5.033042}
}

@article{ZhangSong2013,
  author = {Zhang, X. Z. and Song, Z.},
  title = {Non-{H}ermitian Anisotropic {XY} Model with Intrinsic Rotation-Time-Reversal Symmetry},
  journal = {Phys. Rev. A},
  volume = {87},
  pages = {012114},
  year = {2013},
  doi = {10.1103/PhysRevA.87.012114}
}

@article{ZhangSong2013Geometric,
  author = {Zhang, X. Z. and Song, Z.},
  title = {Geometric Phase and Phase Diagram for a Non-{H}ermitian Quantum {XY} Model},
  journal = {Phys. Rev. A},
  volume = {88},
  pages = {042108},
  year = {2013},
  doi = {10.1103/PhysRevA.88.042108}
}

@article{LiuBatchelor2025,
  author = {Liu, D. C. and Batchelor, Murray T.},
  title = {Characterizing Phase Transitions and Criticality in Non-{H}ermitian Extensions of the {XY} Model},
  journal = {Phys. Rev. B},
  volume = {112},
  pages = {014422},
  year = {2025},
  doi = {10.1103/b55l-7tbc}
}

@article{HenryLiuBatchelor2025,
  author = {Henry, Robert A. and Liu, D. C. and Batchelor, Murray T.},
  title = {Exceptional Point Rings and {$\mathcal{PT}$} Symmetry in the Non-{H}ermitian {XY} Model},
  journal = {AAPPS Bull.},
  volume = {35},
  pages = {29},
  year = {2025},
  doi = {10.1007/s43673-025-00170-w}
}

@misc{LuoMeden2026,
  author = {Luo, Jia-Jia and Meden, Volker},
  title = {Quantum Critical Properties of Non-{H}ermitian {XY} Models with Magnetic Field},
  year = {2026},
  eprint = {2606.07275},
  archivePrefix = {arXiv},
  primaryClass = {quant-ph}
}

@article{LiuXuLi2021,
  author = {Liu, Yu-Guo and Xu, Lu and Li, Zhi},
  title = {Quantum Phase Transition in a Non-{H}ermitian {XY} Spin Chain with Global Complex Transverse Field},
  journal = {J. Phys.: Condens. Matter},
  volume = {33},
  pages = {295401},
  year = {2021},
  doi = {10.1088/1361-648X/ac00dd}
}

@article{PiLu2021,
  author = {Pi, Jinghui and L{\"u}, Rong},
  title = {Phase Diagram and Quantum Criticality of a Non-{H}ermitian {XY} Model with a Complex Transverse Field},
  journal = {J. Phys.: Condens. Matter},
  volume = {33},
  pages = {345601},
  year = {2021},
  doi = {10.1088/1361-648X/ac06ef}
}

@article{LakkarajuSenDe2021,
  author = {Lakkaraju, Leela Ganesh Chandra and Sen(De), Aditi},
  title = {Detection of an Unbroken Phase of a Non-{H}ermitian System via a {H}ermitian Factorization Surface},
  journal = {Phys. Rev. A},
  volume = {104},
  pages = {052222},
  year = {2021},
  doi = {10.1103/PhysRevA.104.052222}
}

@article{AgarwalEtAl2024,
  author = {Das Agarwal, Keshav and Konar, Tanoy Kanti and Lakkaraju, Leela Ganesh Chandra and Sen(De), Aditi},
  title = {Detecting Exceptional Points through Dynamics in Non-{H}ermitian Systems},
  journal = {Phys. Rev. A},
  volume = {110},
  pages = {012226},
  year = {2024},
  doi = {10.1103/PhysRevA.110.012226}
}

@article{WangXu2026,
  author = {Wang, Fei and Liang, Guoying and Zhao, Zecheng and Luo, Lin-Yue and Zhang, Da-Jian and Xu, Bao-Ming},
  title = {Survival of {H}ermitian Criticality in the Non-{H}ermitian Framework},
  journal = {Phys. Rev. B},
  volume = {113},
  pages = {165149},
  year = {2026},
  doi = {10.1103/y34y-32dt}
}

@misc{LiLiuBatchelor2026,
  author = {Li, Yuguan and Liu, D. C. and Batchelor, Murray T.},
  title = {Exact Solution for Non-{H}ermitian Free Fermions: A Case Study of the {XY} Chain},
  year = {2026},
  eprint = {2605.26813},
  archivePrefix = {arXiv}
}

@article{Dzyaloshinsky1958,
  author = {Dzyaloshinsky, I.},
  title = {A Thermodynamic Theory of ``Weak'' Ferromagnetism of Antiferromagnetics},
  journal = {J. Phys. Chem. Solids},
  volume = {4},
  pages = {241--255},
  year = {1958},
  doi = {10.1016/0022-3697(58)90076-3}
}

@article{Moriya1960,
  author = {Moriya, T.},
  title = {Anisotropic Superexchange Interaction and Weak Ferromagnetism},
  journal = {Phys. Rev.},
  volume = {120},
  pages = {91--98},
  year = {1960},
  doi = {10.1103/PhysRev.120.91}
}

@article{KatsuraNagaosaBalatsky2005,
  author = {Katsura, Hosho and Nagaosa, Naoto and Balatsky, Alexander V.},
  title = {Spin Current and Magnetoelectric Effect in Noncollinear Magnets},
  journal = {Phys. Rev. Lett.},
  volume = {95},
  pages = {057205},
  year = {2005},
  doi = {10.1103/PhysRevLett.95.057205}
}

@article{YangLiangCui2023,
  author = {Yang, Hongxin and Liang, Jinghua and Cui, Qirui},
  title = {First-Principles Calculations for {Dzyaloshinskii}--{Moriya} Interaction},
  journal = {Nat. Rev. Phys.},
  volume = {5},
  pages = {43--61},
  year = {2023},
  doi = {10.1038/s42254-022-00529-0}
}

@article{ZhangDM2026,
  author = {Zhang, Panpan and Li, Qinghui and Miao, Chuanzheng and Xu, Yuliang and Yan, Shiwei and Kong, Xiangmu},
  title = {Effects of the {Dzyaloshinsky}--{Moriya} Interaction on Entanglement and Chiral Criticality in a Non-{H}ermitian {XY} System},
  journal = {Phys. Rev. B},
  volume = {113},
  pages = {134433},
  year = {2026},
  doi = {10.1103/8ycw-8lmk}
}

@article{PerkCapel1976,
  author = {Perk, J. H. H. and Capel, H. W.},
  title = {Antisymmetric exchange, canting and spiral structure},
  journal = {Phys. Lett. A},
  volume = {58},
  pages = {115--117},
  year = {1976},
  doi = {10.1016/0375-9601(76)90515-6}
}

@article{ShekhtmanEntinWohlmanAharony1992,
  author = {Shekhtman, L. and Entin-Wohlman, O. and Aharony, A.},
  title = {{Moriya}'s Anisotropic Superexchange Interaction, Frustration, and {Dzyaloshinsky}'s Weak Ferromagnetism},
  journal = {Phys. Rev. Lett.},
  volume = {69},
  pages = {836--839},
  year = {1992},
  doi = {10.1103/PhysRevLett.69.836}
}

@article{Ma2011,
  author = {Ma, Fu-Wu and Liu, Sheng-Xin and Kong, Xiang-Mu},
  title = {Quantum Entanglement and Quantum Phase Transition in the {XY} Model with Staggered {Dzyaloshinskii}--{Moriya} Interaction},
  journal = {Phys. Rev. A},
  volume = {84},
  pages = {042302},
  year = {2011},
  doi = {10.1103/PhysRevA.84.042302}
}

@article{Baran2018,
  author = {Baran, Ostap and Ohanyan, Vadim and Verkholyak, Taras},
  title = {Spin-$1/2$ {XY} Chain Magnetoelectric: Effect of Zigzag Geometry},
  journal = {Phys. Rev. B},
  volume = {98},
  pages = {064415},
  year = {2018},
  doi = {10.1103/PhysRevB.98.064415}
}

@article{Su2022,
  author = {Su, Yao Heng and Liu, D. C. and Wan, Zhongyu and Chen, Ai Min and Cheng, Pengfei},
  title = {Quantum Criticality in Spin-$1/2$ Anisotropic {XY} Model with Staggered {Dzyaloshinskii}--{Moriya} Interaction},
  journal = {Physica A},
  volume = {594},
  pages = {127005},
  year = {2022},
  doi = {10.1016/j.physa.2022.127005}
}

@article{LiuStaggered2024,
  author = {Liu, D. C. and Su, Yao Heng and Jin, Xia and Zhao, Zhijie and Chen, Ai Min},
  title = {Characterizing Quantum Criticality in the Transverse Field {Ising} Model with Staggered {Dzyaloshinskii}--{Moriya} Interaction},
  journal = {J. Magn. Magn. Mater.},
  volume = {590},
  pages = {171659},
  year = {2024},
  doi = {10.1016/j.jmmm.2023.171659}
}

@article{Sun2024,
  author = {Sun, Wen-Yang and Wang, Dong and Shi, Jiadong and He, Juan and Ye, Liu},
  title = {Investigating Quantum Criticality and Multipartite Entanglement in the Anisotropic {XY} Model with Staggered {Dzyaloshinskii}--{Moriya} Interaction},
  journal = {Appl. Phys. B},
  volume = {130},
  pages = {108},
  year = {2024},
  doi = {10.1007/s00340-024-08246-5}
}

@article{Ghosh2010,
  author = {Ghosh, Pijush K.},
  title = {On the Construction of a Pseudo-{H}ermitian Quantum System with a Pre-Determined Metric in the {Hilbert} Space},
  journal = {J. Phys. A: Math. Theor.},
  volume = {43},
  pages = {125203},
  year = {2010},
  doi = {10.1088/1751-8113/43/12/125203}
}

@article{Osterloh2002,
  author = {Osterloh, A. and Amico, Luigi and Falci, G. and Fazio, Rosario},
  title = {Scaling of Entanglement Close to a Quantum Phase Transition},
  journal = {Nature (London)},
  volume = {416},
  pages = {608--610},
  year = {2002},
  doi = {10.1038/416608a}
}

@article{Peschel2003,
  author = {Peschel, Ingo},
  title = {Calculation of Reduced Density Matrices from Correlation Functions},
  journal = {J. Phys. A: Math. Gen.},
  volume = {36},
  pages = {L205--L208},
  year = {2003},
  doi = {10.1088/0305-4470/36/14/101}
}

@article{Calabrese2004,
  author = {Calabrese, Pasquale and Cardy, John},
  title = {Entanglement Entropy and Quantum Field Theory},
  journal = {J. Stat. Mech.},
  pages = {P06002},
  year = {2004},
  doi = {10.1088/1742-5468/2004/06/P06002}
}

@article{PeschelEisler2009,
  author = {Peschel, Ingo and Eisler, Viktor},
  title = {Reduced Density Matrices and Entanglement Entropy in Free Lattice Models},
  journal = {J. Phys. A: Math. Theor.},
  volume = {42},
  pages = {504003},
  year = {2009},
  doi = {10.1088/1751-8113/42/50/504003}
}

@article{EislerPeschel2010,
  author = {Eisler, Viktor and Peschel, Ingo},
  title = {Entanglement in Fermionic Chains with Interface Defects},
  journal = {Ann. Phys. (Berlin)},
  volume = {522},
  number = {9},
  pages = {679--690},
  year = {2010},
  doi = {10.1002/andp.201000055}
}

@article{KarpatCakmakFanchini2014,
  author = {Karpat, G. and {\c{C}}akmak, B. and Fanchini, F. F.},
  title = {Quantum Coherence and Uncertainty in the Anisotropic {XY} Chain},
  journal = {Phys. Rev. B},
  volume = {90},
  pages = {104431},
  year = {2014},
  doi = {10.1103/PhysRevB.90.104431}
}

@article{Lv2022,
  author = {Lv, Cheng-Pu and Li, Yan-Chao and Lin, Hai-Qing},
  title = {Robust Approach to Study the Effect on Quantum Phase Transitions of Various Perturbations at Finite Temperatures},
  journal = {Phys. Rev. B},
  volume = {105},
  pages = {054424},
  year = {2022},
  doi = {10.1103/PhysRevB.105.054424}
}

@article{SuEtAl2024,
  author = {Su, Lin-Lin and Ren, Jun and Ma, Wen-Long and Wang, Z. D. and Bai, Yan-Kui},
  title = {Diagnosing Quantum Phases Using Long-Range Two-Site Quantum Resource Behavior},
  journal = {Phys. Rev. B},
  volume = {110},
  pages = {104101},
  year = {2024},
  doi = {10.1103/PhysRevB.110.104101}
}

@article{LiEtAl2024Multipartite,
  author = {Li, Yan-Chao and Zhou, Yuan-Hang and Zhang, Yuan and Bai, Yan-Kui and Lin, Hai-Qing},
  title = {Multipartite Entanglement Serves as a Faithful Detector for Quantum Phase Transitions},
  journal = {New J. Phys.},
  volume = {26},
  pages = {023031},
  year = {2024},
  doi = {10.1088/1367-2630/ad273a}
}

@article{Ollivier2001,
  author = {Ollivier, Harold and Zurek, Wojciech H.},
  title = {Quantum Discord: A Measure of the Quantumness of Correlations},
  journal = {Phys. Rev. Lett.},
  volume = {88},
  pages = {017901},
  year = {2001},
  doi = {10.1103/PhysRevLett.88.017901}
}

@article{Henderson2001,
  author = {Henderson, L. and Vedral, V.},
  title = {Classical, Quantum and Total Correlations},
  journal = {J. Phys. A: Math. Gen.},
  volume = {34},
  pages = {6899--6905},
  year = {2001},
  doi = {10.1088/0305-4470/34/35/315}
}

@article{Sarandy2009,
  author = {Sarandy, M. S.},
  title = {Classical Correlation and Quantum Discord in Critical Systems},
  journal = {Phys. Rev. A},
  volume = {80},
  pages = {022108},
  year = {2009},
  doi = {10.1103/PhysRevA.80.022108}
}

@article{BaumgratzCramerPlenio2014,
  author = {Baumgratz, T. and Cramer, M. and Plenio, M. B.},
  title = {Quantifying Coherence},
  journal = {Phys. Rev. Lett.},
  volume = {113},
  pages = {140401},
  year = {2014},
  doi = {10.1103/PhysRevLett.113.140401}
}

@article{Girolami2014,
  author = {Girolami, Davide},
  title = {Observable Measure of Quantum Coherence in Finite Dimensional Systems},
  journal = {Phys. Rev. Lett.},
  volume = {113},
  pages = {170401},
  year = {2014},
  doi = {10.1103/PhysRevLett.113.170401}
}

@article{Herviou2019,
  author = {Herviou, Lo{\"i}c and Regnault, Nicolas and Bardarson, Jens H.},
  title = {Entanglement Spectrum and Symmetries in Non-{H}ermitian Fermionic Non-Interacting Models},
  journal = {SciPost Phys.},
  volume = {7},
  pages = {069},
  year = {2019},
  doi = {10.21468/SciPostPhys.7.5.069}
}

@article{Chen2021,
  author = {Chen, Li-Mei and Chen, Shuai A. and Ye, Peng},
  title = {Entanglement, Non-{H}ermiticity, and Duality},
  journal = {SciPost Phys.},
  volume = {11},
  pages = {003},
  year = {2021},
  doi = {10.21468/SciPostPhys.11.1.003}
}

@article{TzengEtAl2021,
  author = {Tzeng, Yu-Chin and Ju, Chia-Yi and Chen, Guang-Yin and Huang, Wen-Min},
  title = {Hunting for the Non-{H}ermitian Exceptional Points with Fidelity Susceptibility},
  journal = {Phys. Rev. Research},
  volume = {3},
  pages = {013015},
  year = {2021},
  doi = {10.1103/PhysRevResearch.3.013015}
}

@article{PiresMacri2021,
  author = {Pires, Diego Paiva and Macr{\`i}, Tommaso},
  title = {Probing Phase Transitions in Non-{H}ermitian Systems with Multiple Quantum Coherences},
  journal = {Phys. Rev. B},
  volume = {104},
  pages = {155141},
  year = {2021},
  doi = {10.1103/PhysRevB.104.155141}
}

@article{Miao2024,
  author = {Miao, Chuanzheng and Li, Yue and Wang, Jie and Zhang, Panpan and Li, Qinghui and Hu, Lizhen and Xu, Yuliang and Kong, Xiangmu},
  title = {Crossover Behavior at an Exceptional Point for Quantum Entanglement and Correlation in a Non-{H}ermitian {XY} Spin System},
  journal = {Phys. Rev. B},
  volume = {110},
  pages = {014403},
  year = {2024},
  doi = {10.1103/PhysRevB.110.014403}
}

@article{ZhangYu2025,
  author = {Zhang, Ling-Feng and Yu, Wing Chi},
  title = {Probing Phase Transitions in Non-{H}ermitian Systems with Quantum Entanglement},
  journal = {Phys. Rev. B},
  volume = {112},
  pages = {144308},
  year = {2025},
  doi = {10.1103/26jz-zmsv}
}

@article{AgarwalEtAl2026,
  author = {Das Agarwal, Keshav and Konar, Tanoy Kanti and Lakkaraju, Leela Ganesh Chandra and Sen(De), Aditi},
  title = {Recognizing Critical Lines Via Entanglement in Non-{H}ermitian Systems},
  journal = {Phys. Rev. A},
  volume = {113},
  pages = {022201},
  year = {2026},
  doi = {10.1103/b5sd-fn57}
}

@article{ArbenzHochstenbach2004,
  author = {Arbenz, Peter and Hochstenbach, Michiel E.},
  title = {A {Jacobi--Davidson} Method for Solving Complex Symmetric Eigenvalue Problems},
  journal = {SIAM J. Sci. Comput.},
  volume = {25},
  number = {5},
  pages = {1655--1673},
  year = {2004},
  doi = {10.1137/S1064827502410992}
}

\end{document}